\documentclass[Afour,sageh,times]{sagej}

\usepackage{moreverb,url}

\usepackage[colorlinks,bookmarksopen,bookmarksnumbered,citecolor=red,urlcolor=red]{hyperref}

\newcommand\BibTeX{{\rmfamily B\kern-.05em \textsc{i\kern-.025em b}\kern-.08em
T\kern-.1667em\lower.7ex\hbox{E}\kern-.125emX}}

\def\volumeyear{2025}
\let\code=\textsf
\let\proglang=\textsf
\newcommand{\SAS}{\textsf{SAS }}

\newcommand{\pkg}[1]{{\fontseries{m}\fontseries{b}\selectfont #1}}

\usepackage{thumbpdf,lmodern}
\usepackage{amsmath}
\usepackage{xcolor}
\usepackage{soul}
\usepackage{caption}
\usepackage{subcaption}

\begin{document}

\runninghead{\pkg{clustra}, Adhikari \& Gerlovin, et.al.}

\title{clustra: A multi-platform k-means clustering algorithm for analysis of longitudinal trajectories in large electronic health records data}

\author{Nimish Adhikari\affilnum{*,1,2},
        Hanna Gerlovin\affilnum{*,1,3},
        George Ostrouchov\affilnum{7},
        Rachel Ehrbar\affilnum{1,2}, 
        Alyssa B. Dufour\affilnum{5},
        Brian R. Ferolito\affilnum{1}, 
        Serkalem Demissie\affilnum{1,2}, 
        Lauren Costa\affilnum{1}, 
        Yuk-Lam Ho\affilnum{1}, 
        Laura Tarko\affilnum{1},
        Edmon Begoli\affilnum{4},
        Kelly Cho\affilnum{1,6},
        and David R. Gagnon\affilnum{1,2} 
}

\affiliation{\affilnum{*} These authors contributed equally\\
\affilnum{1} US Department of Veteran Affairs\\
\affilnum{2} Boston University School of Public Health\\
\affilnum{3} Boston University Chobanian \& Avedisian School of Medicine\\
\affilnum{4} Oak Ridge National Laboratory\\
\affilnum{5} Marcus Institute for Aging Research \\
\affilnum{6} Harvard Medical School \\
\affilnum{7} University of Tennessee, Business Analytics and Statistics}

\corrauth{Hanna Gerlovin, U.S. Department of Veterans Affairs,
  2 Avenue de Lafayette Attn: VA,
  Boston, Massachusetts 02111}

\email{hanna.gerlovin@va.gov}
\begin{abstract}
  \textit{Background and Objective}: Variables collected over time, or longitudinally, such as biologic measurements in electronic health records data, are not simple to summarize with a single time-point, and thus can be more holistically conceptualized as trajectories over time. Cluster analysis with longitudinal data further allows for clinical representation of groups of subjects with similar trajectories and identification of unique characteristics, or phenotypes, that can be investigated as risk factors or disease outcomes. Some of the challenges in estimating these clustered trajectories lie in the handling of observations at inconsistent time intervals and the usability of algorithms across programming languages.
  
  \textit{Methods}: We propose longitudinal trajectory clustering using a k-means algorithm with thin-plate regression splines, implemented across multiple platforms, the \proglang{R} package \pkg{clustra} and corresponding \SAS macros. The \SAS macros accommodate flexible clustering approaches, and also include visualization of the clusters, and silhouette plots for diagnostic evaluation of the appropriate cluster number. The \proglang{R} package, designed in parallel, has similar functionality, with additional multi-core processing and Rand-index-based diagnostics. 
  
  \textit{Results}: The package and macros achieve comparable results when applied to an example of simulated blood pressure measurements based on real data from Veterans Affairs Healthcare recipients who were initiated on anti-hypertensive medication. 

  \textit{Conclusion}: The \proglang{R} package \pkg{clustra} and the \SAS macros integrate a K-means clustering algorithm for longitudinal trajectories that operates with large electronic health record data. The implementations provide comparable results in both platforms, satisfying the needs of investigators familiar with, or constrained by access to, one or the other platform.

\end{abstract}


\maketitle

\section{Copyright Statement}

This manuscript has been authored in part by UT-Battelle, LLC, under contract DE-AC05-00OR22725 with the US Department of Energy (DOE). The publisher acknowledges the US government license to provide public access under the DOE Public Access Plan (\url{http://energy.gov/downloads/doe-public-access-plan})
\begin{acks}
This research is based on data from the Million Veteran Program, Office of Research and Development, Veterans Health Administration, and was supported by award No.~MVP000. HG is supported by the VA Office of Research and Development Cooperative Studies Program (CSP) award (\#2032). This research used resources from the Knowledge Discovery Infrastructure (KDI) at Oak Ridge National Laboratory, which is supported by the Office of Science of the US Department of Energy under Contract No. DE-AC05-00OR22725.

This manuscript has been in part co-authored by UT-Battelle, LLC under a joint program with the Department of Veterans Affairs under the Million Veteran Project Computational Health Analytics for Medical Precision to Improve Outcomes Now (MVP-CHAMPION).

This publication does not represent the views of the U.S. Department of Veterans Affairs or the United States Government.

\end{acks}

\section{Introduction}
In electronic health records (EHR) and administrative healthcare databases, biologic measurements are collected over long periods for many patients. Single measurements of biomarkers or phenotype can provide valuable risk information, but repeated measurements offer deeper insights into individual changes over time. Cluster analysis of these longitudinal trajectories can group subjects with similar trajectories, revealing  unique phenotypes that may serve as risk factors or outcomes themselves.

Several methods and algorithms have been proposed for classifying longitudinal data into groups \citep{Genolini2011,Genolini2015,Stoitsas2022,Teuling2023, BUCHIN2019}, typically assuming regular observation intervals. Unlike prospective cohort studies, medical records often contain irregularly spaced data due to varying clinical presentations and administrative touch-points, complicating pattern recognition. \cite{Luong2017} proposed a k-means approach allowing arbitrary time points, but provided no software implementation. Challenges in clustering algorithms also include software availability, as different institutions prefer specific platforms like SAS or R. For example, \cite{Genolini2016} developed a shape-based approach using Fr\'echet distance in the \proglang{R} package \pkg{kmlShape}, now archived in CRAN, highlighting the need for flexible solutions aligned with clinical judgment.  

The manuscript outlines a k-means clustering framework for identifying the latent clusters within longitudinal data, implemented in both \SAS and \proglang{R}, and address platform-specific considerations (\hyperref[sec:methods]{Methods}). Using simulated blood pressure data from US Veterans on anti-hypertensive therapy, we demonstrate the utility of the multi-platform approach with the \pkg{clustra} package in \proglang{R} and corresponding \SAS macros, ensuring consistent, reproducible results across both environments (\hyperref[sec:dataex]{Results}).

\section{Methods} \label{sec:methods}
\subsection{K-means Clustering with Smoothing Splines}
Clustering by K-means follows an expectation maximization (EM) algorithm to identify the optimal group assignment per subject \citep{Genolini2015,Luong2017}. Subjects are initially assigned randomly into \$K\$ clusters. Cluster centers are then defined and distances from observations to these cluster centers are calculated. Subjects are then moved to the cluster with the nearest cluster center. The cluster center is then recalculated and movement continues. This iterative process continues until few or no subjects change cluster classification between steps, depending on the pre-specified stopping criteria. 

The traditional k-means approach for longitudinal data attempts to reduce dimensionality from $n$ trajectories (for $n$ subjects) to $k$ informative clusters. The distance between a single trajectory and the average for a cluster can be calculated relative to the number of observations, thereby circumventing problems with missing and irregular data. In particular, a generalized additive model (GAM) \citep{Hastie1995} can be fit using thin plate regression splines with a penalized number of knots to achieve smoothing \citep{Wood2003}. This approach has been implemented in \proglang{R} using the \pkg{mgcv} package \citep{Wood2003,Wood2017} and in \SAS with \pkg{PROC GAMPL}. Although the implementations differ, they reference the same underlying methodology in \cite{Wood2003}.

Generalized additive models (GAMs) are used to model the non-linear relationship between an outcome and a set of predictors. We are interested in predictors that are smooth nonlinear functions of time to model a trajectory. That is,
\begin{equation*}
    y_i = f(t_i) + \epsilon_i \quad\mbox{for}\quad i = 1, \ldots, n,
\end{equation*}
where $y_i$ are the responses, $t_i$ are the observed time points, and $f()$ is a smooth nonlinear function. When smoothness is defined by a second derivative of $f()$ penalty, the optimization
\begin{equation}
    \sum_{i=1}^n (y_i - f(x_i))^2 + \lambda f''(x_i)^2
    \label{eq:pr}
\end{equation}
attains minimum for $f()$ as a cubic spline \citep[Section 5.4]{Hastie1995}, which is a special case of thin plate spline (TPS) \citep{Wood2003}. TPS is a smoothing spline, where the smoothing parameter $\lambda\in [0, \infty)$ is chosen by cross-validation. The advantage of using a penalized model is in the ability for each cluster to have different degrees of freedom used in the modeling as they will typically have different numbers of observations.

A cluster “center” is defined by the resulting predicted values of the TPS estimated from the combined observations of subjects in that cluster. More formally, we have subjects $i = 1, \ldots, I$, where subject $i$ has $n_i$ responses $\{y_{i1}, \ldots, y_{in_i} \}$ at times $\{t_{i1}, \ldots, t_{in_i} \}$. Let cluster $k$ contain $m_k$ subjects indexed by $\{i_1,i_2,\ldots,i_{m_k} \} \subset \{1,\ldots,I\}$. Note that the $i_j$'s are different for each $k$ but we omit the $k$ index to simplify notation. Cluster $k$ has a total of
\begin{equation*}
    s_k = \sum_{j=1}^{m_k} n_{(i_j)}
\end{equation*}
responses among the $m_k$ subjects. All $s_k$ time points and corresponding responses are used to estimate one TPS $\hat{f_k}()$, as the cluster $k$ "center."

Denote by $D_{ik}$ the distance of subject $i$ to cluster $k$ center and define it as the mean squared distance at that subject's observed times. That is,
\begin{equation}
    D_{ik} = \frac{1}{n_i} \sum_{j=1}^{n_i} \left( y_{ij} - \hat{f_k}(t_{ij}) \right)^2.
    \label{eq:dist}
\end{equation}
The minimum value over $k$ of $D_{ik}$ for subject $i$ indicates the closest cluster for that subject during that iteration.

\subsection{K-means algorithm steps}
Given the distance metric (\ref{eq:dist}), the K-means algorithm is as follows:
\begin{enumerate}
    \item Initially, randomly assign subjects to one of $k = 1,\ldots,K$ clusters.
    \item \label{it:fit} Fit a thin plate regression spline $\hat{f_k}()$ for each cluster, using subjects currently assigned to the cluster.
    \item Compute the distances $D_{ik}$ (\ref{eq:dist}) of each subject $i$ to the estimated cluster centers $\hat{f_k}()$, $k = 1,\ldots, K$, from step \ref{it:fit}.
    \item Find $\min_k D_{ik}$ for each subject $i$ and re-assign it to that “closest” cluster.
    \item Calculate the number of subjects that switched groups during this iteration.
    \item If the maximum number of iterations has not been met and the proportion of subjects changing groups is greater than the convergence criteria, increment the iteration count by 1 and return to step 2.
    \item Once the maximum number of iterations has been met or the proportion of subjects switching groups has gone below the threshold, stop iterating.
\end{enumerate}
Note that at each iteration, it is possible for a particular group to have no remaining subjects. In this case, the cluster is dropped and the algorithm collapses to $K-1$ (or fewer) clusters. It is also possible that the remaining subjects in a cluster jointly do not have enough time points to start the cross-validation estimation of (\ref{eq:pr}), in which case the subjects are assigned to the next closest center and the cluster is also dropped.

\subsection{Selecting the Appropriate Number of Clusters}
Many approaches have been proposed for selecting the correct number of clusters or groups in a K-means algorithm. To say that this question has not yet been answered discounts the tremendous number of possible approaches that have been proposed \citep{Hofmeyr2020,Liu2010,tsclust2014,clValid2008}. For the purposes of our algorithm, we focused on several accepted strategies: the Rand Index \citep{Rand1971}; silhouette plots \citep{Rousseeuw1987}; and using the results of a two stage clustering approach with hierarchical clustering techniques like dendrograms. All three allow the user a graphical assessment of the number of clusters selected. While many other options exist, these can be implemented using the output data and predicted values from the final iteration of the algorithms presented here.

\subsubsection{Rand Index}
Rand index \citep{Rand1971} evaluates similarity between two clustering assignments of a set of subjects. It is an objective measure that counts how many pairs of subjects (among all possible pairs) are grouped together or apart in the two clustering assignments. Given two sets of cluster assignments $a$ and $b$, the Rand index varies from 0 to 1 and is given by
\begin{equation*}
    R_{ab} = \frac{n_{aa} + n_{bb}}{\frac{1}{2}n(n-1)},
\end{equation*}
where $n_{xx}$ in the numerator is the number of pairs in the same cluster according to cluster assignment $x$ and the denominator is the number of possible pairs among $n$ units. There are a few ways to adjust for random agreement, where we use the most popular Adjusted Rand Index (ARI) \citep{Hubert1985} that adjusts for expected random agreement
\begin{equation*}
    AR_{ab} = \frac{R_{ab} - E[R_{ab}]}{1 - E[R_{ab}]}.
\end{equation*}
The ARI implementation in \pkg{MixSim} \citep{mixsim2012} is used by the \pkg{clustra} \proglang{R} package (ARI is currently not available in the \SAS macros).

The ARI plot (see Fig.~\ref{fig:clustra_rand} for an example)
compares clustering results of several random starts and several $k$. Typically, the highest value of $k$ for which the index is consistently high between random starts is taken as the recommended number of clusters \citep{Chen2013APE}.

\subsubsection{Silhouette Plot}
A silhouette plot is a graphical presentation of cluster separability. It compares subject distances to their assigned nearest cluster to that of their next-nearest cluster. The silhouette ranges from -1 to +1, where a value closer to +1 indicates that the object strongly matched to its own cluster and poorly matched to the neighboring clusters. If most of the objects have a high value closer to +1, then our clustering configuration is appropriate, but if many points have low or negative values then we might have too many or too few clusters \citep{Rousseeuw1987}.

\subsubsection{Dendrogram}
A dendrogram is a type of tree diagram that shows hierarchical clustering results for different numbers of clusters. The horizontal line in the diagram is called a clade and every clade has one or more leaves. The clades are arranged according to how similar they are. Clades that are closer in height are more similar, and clades that are further are dissimilar. Looking at the dendrogram can give us an idea about the appropriate number of clusters for the data based on how similar/dissimilar certain clades are. An example dendrogram is provided in section 4 alongside the data example.
\subsection{Developing the packages}
\label{sec:packages}

Package and macro development relies on native components already implemented on each platform. This presents some challenges because the components may have some differences in implementation even when the same methodology is used. For example, both \SAS and \proglang{R} use the Mersenne Twister algorithm \citep{Matsumoto1998MersenneTA} for random number generation. In \proglang{R}, the algorithm produces
\begin{verbatim}
> RNGkind()
[1] "Mersenne-Twister" "Inversion"    
"Rejection"       
> set.seed(987654321)
> runif(6)
 [1] 0.36888095 0.80854246 0.54802340
 0.17993140 0.65435477 0.12500433
 
\end{verbatim}
Invoking the RNG with the same seed in \SAS produces:
\begin{verbatim}
data A;
	call streaminit(987654321);
	do i = 1 to 6;
		u = rand("Uniform");
		output;
	end;
run;

0.27419 0.14882 0.22030 0.48942 0.26644 0.93453 
    
\end{verbatim}
The different sequence is likely due to implementation choices, which include different ways of initializing from the same seed. Even the original authors have later suggested different initializations. As a result, our goal is to produce equivalent but not necessarily identical results.

\subsection{SAS Macros}
\label{sec:sas}

The \SAS macros are self-contained and were originally developed by Gagnon and have been used for clustering frailty trajectories before death \citep{Ward2021}. They are contained in a single \SAS program that should be executed at the beginning to initialize them in the system. This program also contains \SAS code to create a style definition and default graphics attributes. 

There are multiple \SAS macros because of the utility of certain sections of code for additional uses (Table \ref{SAStable}), e.g., the \code{\%TRAJPLOT} macro can be used for additional graphics production. While \code{PROC GAMPL} is the default procedure for producing predicted splines, the \code{\%CLUSSMOOTH} macro, which is called by the \code{\%TRAJLOOP} macro provides several alternative variants of \code{PROC GAMPL} as well as the \code{GLIMMIX, TPSPLINE, GAM and TRANSREG} procedures. \code{PROC GAMPL} is preferred, as it implements penalized splines and also is a "high performance analytic procedure" which is capable of using multiple threads and distributed computing. \citep{GAMPLfeatures} Multi-threading options can be easily modified by the user in the \code{PERFORMANCE} statements in the  \code{\%CLUSSMOOTH} macro. Full details of the fitting functions available in the \SAS Macros are provided in Appendix \ref{appendix:SASMacro}. Parameter options for each Macro can be found in 
\href{https://github.com/MVP-CHAMPION/clustra-SAS}{MVP\_CHAMPION/clustra-SAS}.
\begin{table}[h]
    \small\sf\centering
\begin{tabular}{lp{2in}}
  \hline
  \bf SAS Macro & \bf Function  \\\hline
\%TRAJSETUP & The first macro to execute. Initializes the clustering process, assign initial random   clusters. Execute before calling \%TRAJLOOP for the first time   \\\hline

\%TRAJLOOP  & Execute the  clustering algorithm. May be repeated if more iterations are desired  \\\hline

\%CLUSSMOOTH & Defines the   form of smoothing used in \%TRAJLOOP and other macros.  The METHOD is set in the \%TRAJSETUP macro   for use in \%TRAJLOOP, with GAMPL with default optimization as the default method \\\hline
\%TRAJPLOT  & Macro for   printing cluster analysis results.  May   be used for additional copies of trajectory output without re-running the   \%TRAJLOOP macro.  This macro is called   within \%TRAJLOOP.   \\\hline
\%SILHOUETTE & Creates   silhouette plots.  This macro uses   output from the \%TRAJLOOP macro  \\\hline
\%TRAJFIT   & This macro   will run the \%TRAJSETUP \& \%TRAJLOOP macros multiple times with different   numbers of clusters, extracting the AIC statistic from each run for   comparisons \\\hline
\%TRAJHCLUS & For 2-stage   clustering, do an initial K-means clustering using \%TRAJSETUP and \%TRAJLOOP   with more than the desired final number of clusters. This macro uses output   from \%TRAJLOOP and performs hierarchical    clustering on the outputted clusters. \\\hline
\%GRAPHSET  & A utility   macro for graphics output options.    Should be run before macros that produce graphics output  \\\hline
Other \SAS   code & The program   contains PROC TEMPLATE code that defines a SPLINECURVE style for graphics and   also creates a graphics attributes data set for color and other graphics options                                
 \\\hline
\end{tabular}
\captionof{table}{SAS Macro function descriptions}\label{SAStable}
\end{table}

\subsection{Development of the R package}
\label{sec:rpack}

As part of the MVP CHAMPION interagency collaboration between the Department of Energy’s Oak Ridge National Laboratory (ORNL) and Department of Veteran’s Affairs Million Veteran Program (MVP) Data Core, the algorithm was shared and implemented using the most appropriate \proglang{R} functions by Ostrouchov into the \pkg{clustra} \proglang{R} package. An outer loop of an EM algorithm uses the \pkg{mgcv} package for TPS fitting of data that is managed with the \pkg{data.table} package in addition to some custom functions that assign subjects to clusters based on nearest cluster center and test convergence criteria.

We use the \code{bam()} function from \pkg{mgcv} for its ability to work with very large numbers of observations by its use of discretization of covariate values. In addition, there are opportunities for parallelization built in to \pkg{data.table} and \pkg{mgcv}. We also add parallelization over $K$ with \code{mclapply()} from the \pkg{parallel} package for fitting and prediction within the EM iterations. \pkg{data.table} controls core use with its \code{setDTthreads()} function. The \code{bam()} function in \pkg{mgcv} uses the parameter \code{nthreads} to control core use in its \code{OpenMP} compiled \code{C} code. In addition, we can also provide an optimized \code{OpenMP} compiled \pkg{OpenBLAS} library. With all these potential demands for cores, careful management is necessary.

Running a large number of benchmarks with data containing roughly 30,000 subjects, each with an average of 25 observations, on a high core count cluster node revealed that none of the built-in parallelizations produce much speedup. This is likely because our use case of fitting a TPS with a single covariate, time, does not result in a large-enough number of TPS columns. Some improvement of about 10\% is observed when a single threaded \pkg{OpenBLAS} is substituted for the default \pkg{NETLIB} version. Only the parallelization over $K$ clusters within the main EM loop produces reliable speedups with up to $K$ cores.

The Rand Index evaluation of $K$ is only available in the \pkg{clustra} \proglang{R} code and it provides further parallelization opportunity through the need to run several random starts for each $K$. Here too, reliable speedups are obtained up to the full node with 32 cores, when requesting 10 replicates for $K$ = 2, 3, and 4.

The package vignette, \code{clustra\_vignette.Rmd}, runs through data generation and then exercises all functions and adds various plots to show the results. The \code{clustra()} function is at the top level. It initializes the clusters and uses the \code{trajectories()} function to perform the k-means EM iteration and check stopping criteria. We opted to use only one parallelization parameter \code{mccores} in the \code{clustra()} function for parallelization over $K$ and set all the built-in parallelizations to 1 core.

Among additional functions, \code{clustra\_sil()} either runs \code{clustra()} or takes in a previous  \code{clustra} object and prepares silhouette plot data (the plot code is in the vignette), and the \code{clustra\_rand()} function performs a \code{trajectories()} run for multiple numbers of clusters with random restarts and returns a dataframe with a Rand Index comparison between all pairs of clusters. A plot from the resulting dataframe can be produced with the \code{rand\_plot()} function.

\subsection{How do SAS and R compare?}

The \SAS macros use variations of \code{PROC GAMPL} and the \proglang{R} package uses \code{mgcv::bam} for it's approach towards regression splines, and though they are based on the same methods, there are slight differences of implementations between the two programming languages. The \SAS macro \code{\%TRAJPLOT} is capable of producing longitudinal cluster plots and \pkg{clustra} has the function \code{plot\_smooths()} with a similar capability and the ability to include varying amounts of data in the same plot. The \pkg{clustra} also contains functions \code{clustra\_rand} and \code{rand\_plot} that provide a Rand index, as well as a matrix plot, that can be used to determine the appropriate number of clusters. The \SAS macros do not have Rand index capabilities, however the output can be read into R using the \pkg{haven} package and compared with the \proglang{R} output using the \code{MixSim::RandIndex} function in R. A complete table of comparable functions can be found in Appendix \ref{app:functions}, and more details regarding the application of each code are demonstrated in Section \ref{sec:dataex}. The comparison between the output of the \SAS and \proglang{R} packages for the data example can also be found in Appendix \ref{app:cluster_comp}.

\section{Results} \label{sec:dataex}

The following section illustrates the functions and commands in the package and macro on a example data set, as well as provide methods to select an appropriate number of clusters, which can be tested against the true cluster size using the adjusted rand index. The detailed code and data to run this example can be found in repositories \pkg{clustra} and \pkg{clustra-SAS} of the \href{https://github.com/MVP-CHAMPION}{MVP\_CHAMPION} organization.

\subsection{Example data source}
\label{sec:datasource}
The US Veteran’s Affairs (VA) Healthcare System is the nation’s largest EHR system, housing administrative data for over 24 million Veterans enrolled since 1998. Clinical encounters, laboratory measures, and pharmacy data are collected across over 114 medical facilities and even more community-based outpatient clinics. While not all Veterans who are enrolled at the VA receive consistent care through VA-providers, there are roughly 6 million individuals with at least one primary care physician visit per year.

For this study, we collected data on veterans initiating anti-hypertensive therapy. Time=0 was defined as the date of the first prescription of an anti-hypertensive medication. We collected all measures of systolic blood pressure [SBP] from one year prior to two years post time zero (or -365 to 730 days). To account for data sparsity, we required all individuals to have at least one SBP measurement in the year before initiation, and a minimum of 3 SBP measures after time zero. Individuals had to have survived through the two-year follow-up period, though no other restrictions were placed on the cohort of Veterans.

Using this population of subjects who were initiating new anti-hypertensive therapy at time=0, it is expected that all subjects will have some relatively high SBP measurements prior to starting therapy, have a relatively fast decline in SBP immediately after starting therapy, and also have some changes after this drop, either maintaining the decline, further declining or slowly rising. It is hypothesized that there are multiple groups of subjects in this population that will have various patterns of SBP trajectories over this time period. While cluster analysis will identify clusters of subjects with similar SBP patterns over time under any conditions, the utility of this approach is seen if these clusters provide clinically meaningful groups where these groups differ on measurable outcomes, such as mortality.

\subsection{Simulated data}
\label{sec:simdata}
The data used in the following example was simulated based off of the real data from the VA Healthcare system. Details about the simulation process are not within the scope of this paper, though a brief description follows. We used real data on Veterans with blood pressure measurements over time in the VHA database, who had initiated anti-hypertensive therapy, as described previously. The real data was clustered into 5 groups using the \SAS macros, with the output parameters from those groups retained as the “true” trajectories. These parameters were then used to simulate a group, time and response variables for 80,000 individuals, with variation built into the number of observations and values of the responses, to allow the data to appear more realistic or “messy”. For the purpose of this example the true underlying structure is with 5 clusters. This simulated data set of 1,353,910 rows and 4 columns is available as a gzipped CSV file in the GitHub \href{https://github.com/MVP-CHAMPION/clustra-SAS}{MVP\_CHAMPION/clustra-SAS} repository under its \textsf{bp\_data} directory. A subset of this data, with 10,000 randomly sampled individuals, is included in the \pkg{clustra} package as \code{data(bp10k)}.

\subsection{Application}

The \SAS Macro and \proglang{R} package will be used to cluster the simulated data described above, into 2, 5 and 10 clusters. Additional comparisons for different options in the \code{clustra} function and the \SAS macro call will also be performed. The output from the application of the \proglang{R} package and the \SAS macro will be compared to the true group assignments of the simulated data using Adjusted Rand Index.

\subsubsection{R implementation of the Data example}
For the \proglang{R} implementation of this example, some other packages alongside \pkg{clustra} are required. These are \pkg{mgcv}, \pkg{data.table}, and \pkg{MixSim}. To run the package vignettes, additional packages \pkg{haven}, \pkg{knitr}, and \pkg{rmarkdown} are needed. They can be installed into \proglang{R} using standard approaches, including the \code{install.packages()} command or your RStudio Packages interface. Specifically, the vignette \code{clustra\_bp\_vignette.Rmd} can reproduce the results in this example. In the code that is shown here, we omit repeating \code{set.seed(12345)} that is used in the vignette for reproducibility.

To run \pkg{clustra} functions, you only need to load \code{library(clustra)} as functions from the other packages are referenced via the \pkg{clustra} namespace and need not be loaded in their entirety.
\begin{verbatim}
library(clustra)
\end{verbatim}

The 10,000 individuals sample of the simulated dataset is available in the package as \code{bp10k}. We just call it \code{data} so we can continue in a generic way.
\begin{verbatim}
data = bp10k
\end{verbatim}
Alternatively, the full data set with 80,000 individuals can be read in with
\begin{verbatim}
repo = "https://github.com/MVP-CHAMPION/"
repo_sas = paste0(repo, "clustra-SAS/raw/main/")
url = paste0(repo_sas, 
    "bp_data/simulated_data_27June2023.csv.gz")
data = data.table::fread(url)
\end{verbatim}
We are following components of \code{clustra\_bp\_vignette.Rmd}, where the above data alternatives and the running of various chunks are controlled with parameters \code{chunk}, \code{full\_data}, and \code{mc}, resulting in vignette run times ranging from about a minute to several hours. The results presented in what follows use the full data set of 80,000 individuals with a total of 1,353,910 observations. 

The \code{clustra()} function from our package can be used as follows:
\begin{verbatim}
cl5 = clustra(data, k = 5, maxdf = 30, 
    conv = c(20, 0.5), mccores = mc, 
    verbose = TRUE)
\end{verbatim}

We set the number of clusters, \code{k = 5}, the spline max degrees of freedom to 30, the maximum number of EM iterations to 20, and the minimum percentage of subjects changing groups to continue iterations as 0.5. The \code{mccores} parameter sets the number of cores to use in the code components, and should be set to the number of clusters for optimal performance. The verbose parameter provides full information regarding the iterations, as well as the runtime in hours.

\begin{figure}
     \centering
     \begin{subfigure}[b]{0.48\textwidth}
         \centering
         \includegraphics[width=\textwidth]{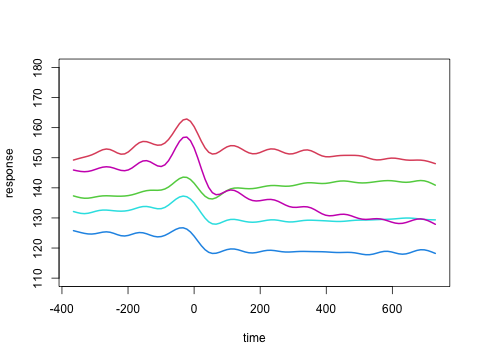}
         \caption{R}
         \label{fig:R_cl5_sub}
     \end{subfigure}
     \hfill
     \begin{subfigure}[b]{0.48\textwidth}
         \centering
         \includegraphics[width=\textwidth]{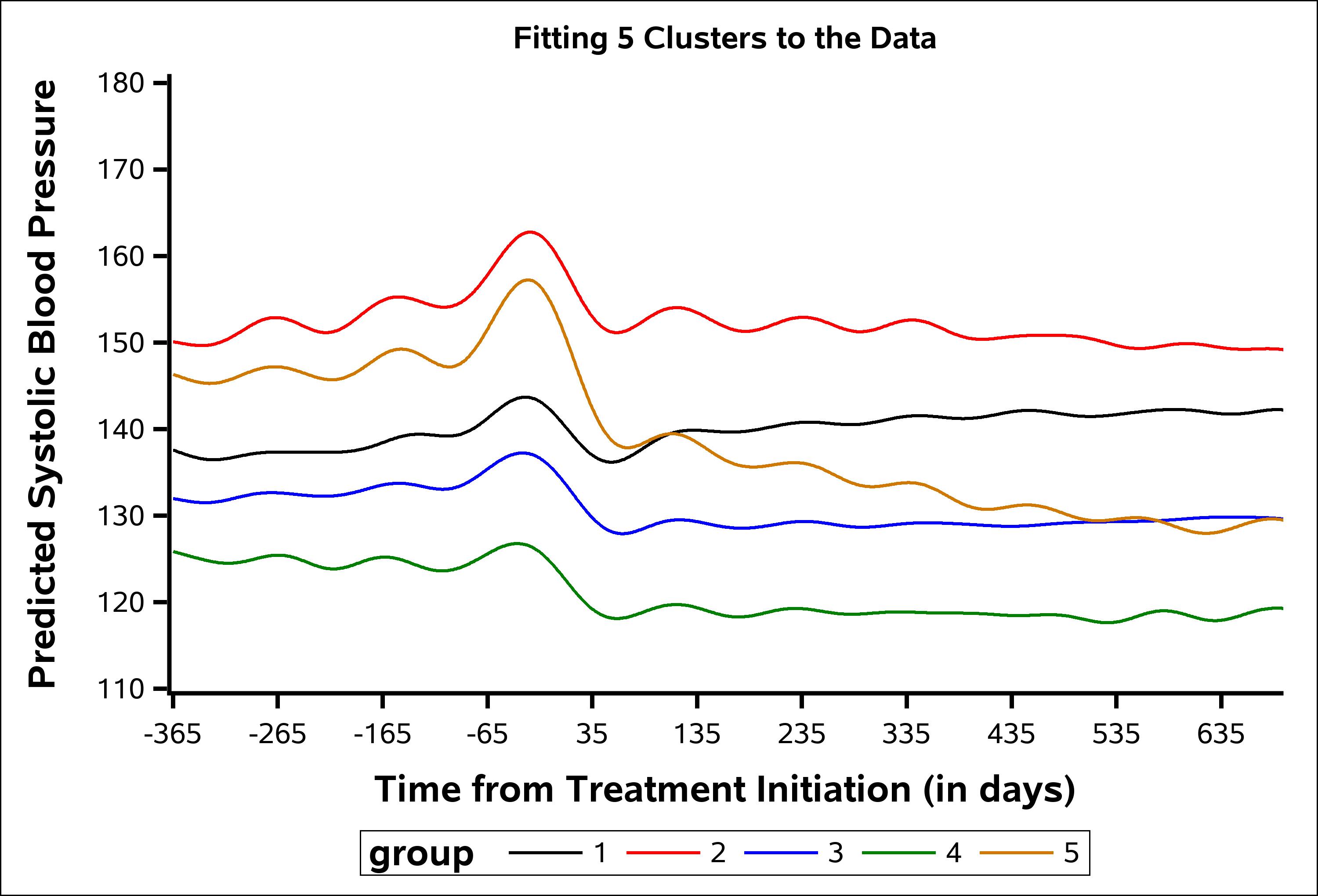}
         \caption{SAS}
         \label{fig:SAS_cl5_sub}
     \end{subfigure}
     \hfill
        \caption{Clustering the data into 5 trajectories for the two implementations}
        \label{fig:cl5}
\end{figure}

Here, the EM algorithm converges after 8 iterations and the output from the function, \code{cl5}, is a list with components defined in the package documentation. The \code{plot\_smooths} function can then be used to plot the estimated cluster trajectory smooths (cluster means). The parameters of this function are the data, the \code{tps} component of the \code{clustra} output, \code{max.data} limits the number of data points to plot, and \code{ylim} optionally controls the y-axis of the plot. Using this function we plot the 5 cluster results in Figure \ref{fig:R_cl5_sub}.
\begin{verbatim}
plot_smooths(data, fits = cl5$tps, max.data = 0,
    ylim = c(110, 180))
\end{verbatim}

The rand index for comparing the \code{cl5} output with the true groups is given by
\begin{verbatim}
MixSim::RandIndex(cl5$data_group, 
    data[, true_group])$AR
> 0.7026215
\end{verbatim}

\begin{figure}
     \centering
     \begin{subfigure}[b]{0.48\textwidth}
         \centering
         \includegraphics[width=1.05\textwidth]{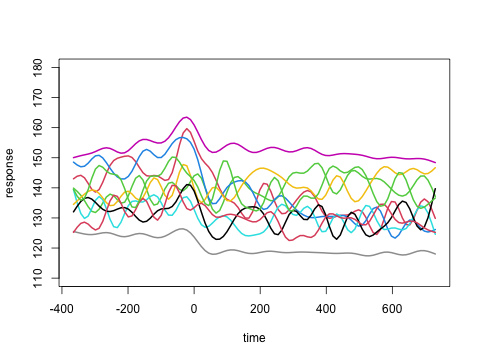}
         \caption{R}
         \label{fig:R_cl10_sub}
     \end{subfigure}
     \hfill
     \begin{subfigure}[b]{0.48\textwidth}
         \centering
         \includegraphics[width=\textwidth]{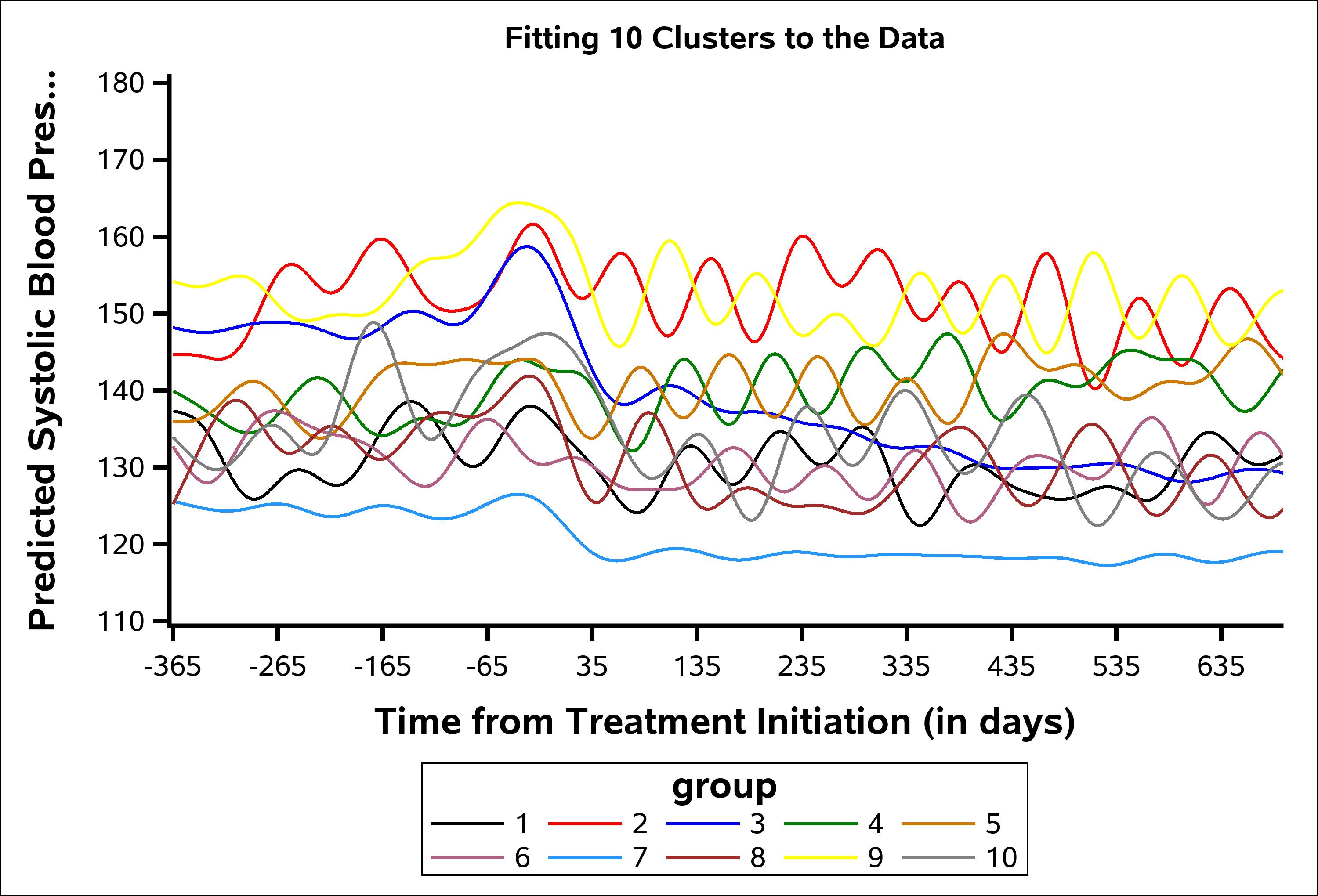}
         \caption{SAS}
         \label{fig:SAS_cl10_sub}
     \end{subfigure}
     \hfill
        \caption{Clustering the data into 10 trajectories for the two implementations}
        \label{fig:cl10}
\end{figure}

\begin{figure}
     \centering
     \begin{subfigure}[b]{0.48\textwidth}
         \centering
         \includegraphics[width=\textwidth]{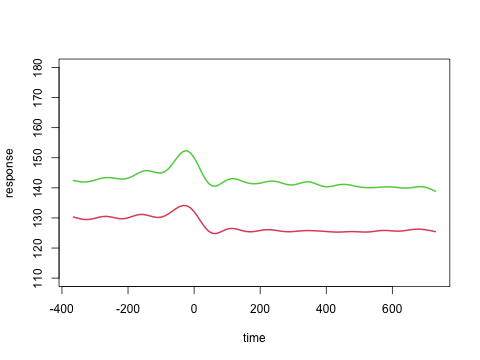}
         \caption{R}
         \label{fig:R_cl2_sub}
     \end{subfigure}
     \hfill
     \begin{subfigure}[b]{0.48\textwidth}
         \centering
         \includegraphics[width=\textwidth]{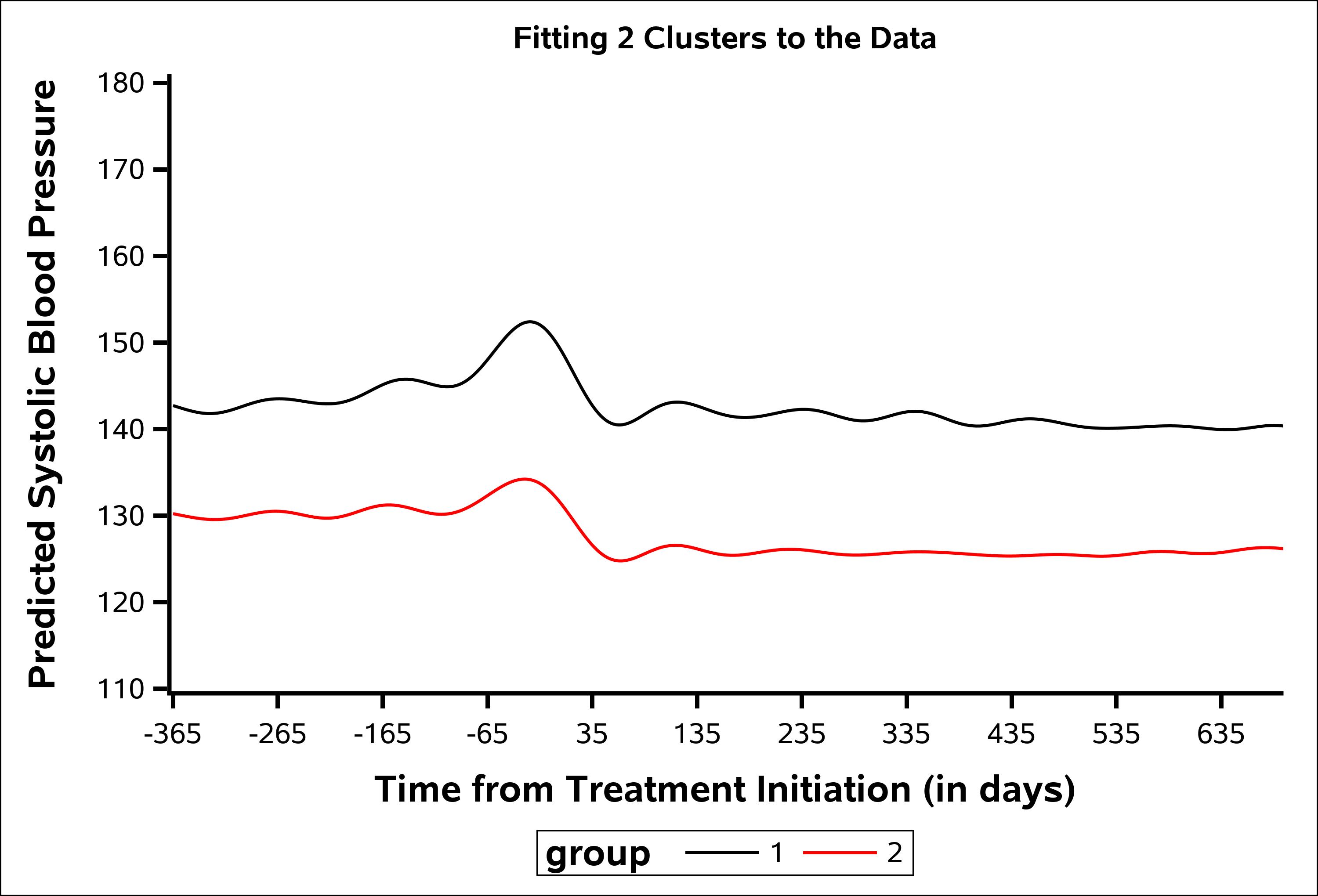}
         \caption{SAS}
         \label{fig:SAS_cl2_sub}
     \end{subfigure}
     \hfill
        \caption{Clustering the data into 2 trajectories for the two implementations}
        \label{fig:cl2}
\end{figure}

Next, we request 10 and 2 clusters with max degrees of freedom as 30, the maximum number of EM iterations as 50, and minimum iteration percentage as 0.5 and plot the trajectories and get an adjusted rand index with respect to the true groups using the function defined above. The results can be seen in Figures \ref{fig:R_cl10_sub} and \ref{fig:R_cl2_sub}. 
\begin{verbatim}
cl10 = clustra(data, k = 10, maxdf = 30, 
    conv = c(50, 0.5), mccores = mc)
plot_smooths(data, cl10$tps, max.data = 0, 
    ylim = c(110, 180))
MixSim::RandIndex(cl10$data_group, 
    data[, true_group])$AR
> 0.4131535

cl2 = clustra(data, k = 2, maxdf = 30, 
    conv = c(20,0.5), mccores=mc)
plot_smooths(data, cl2$tps, max.data = 0, 
    ylim = c(110, 180))
MixSim::RandIndex(cl2$data_group,
    data[, true_group])$AR
> 0.342113
\end{verbatim}

\begin{figure}[htbp]
     \centering
     \begin{subfigure}[b]{0.48\textwidth}
         \centering
         \includegraphics[width=\textwidth]{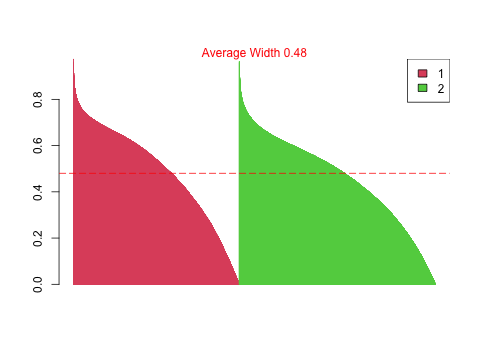}
         \caption{R}
         \label{fig:R_cl2_sil}
     \end{subfigure}
     \hfill
     \begin{subfigure}[b]{0.48\textwidth}
         \centering
         \includegraphics[width=\textwidth]{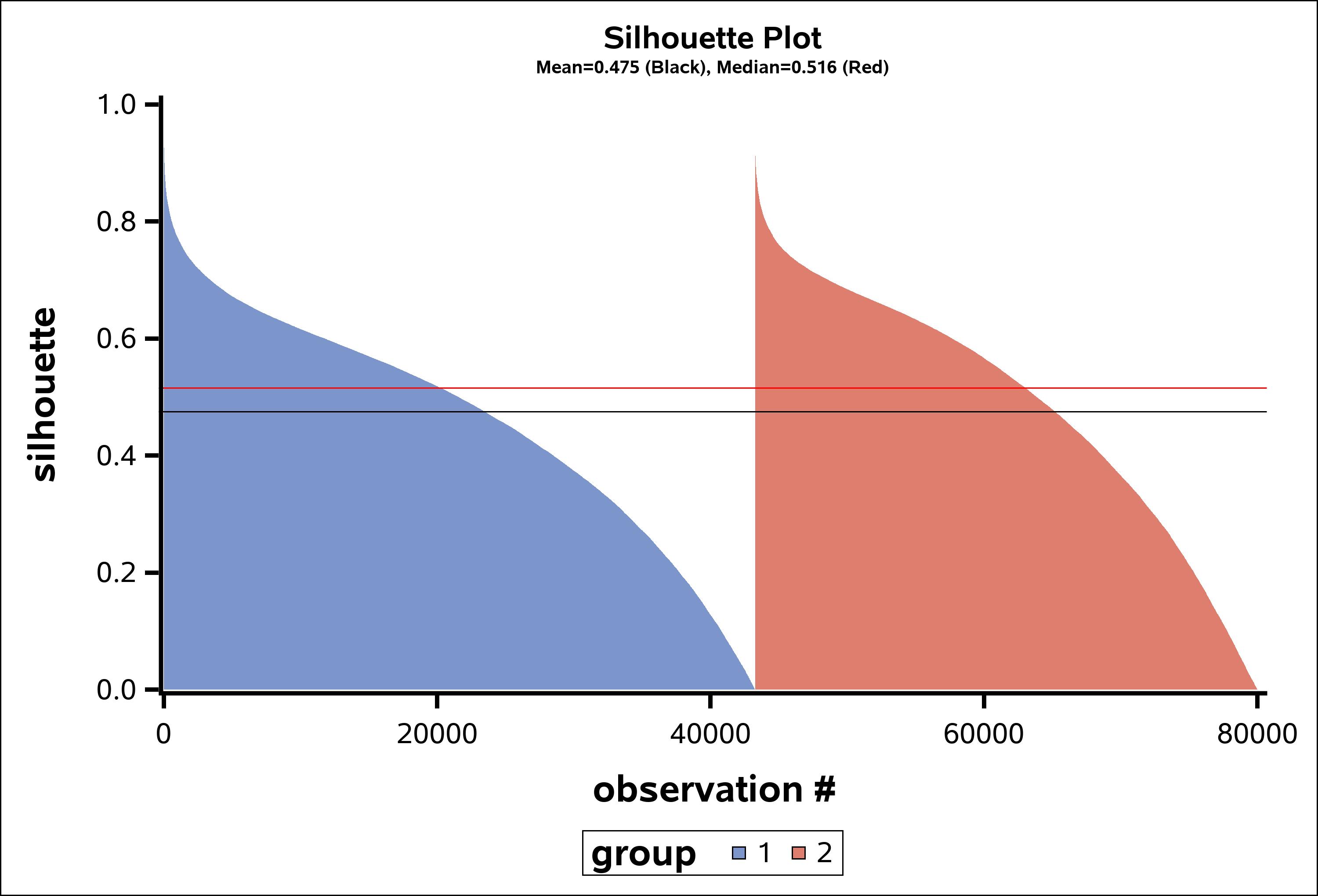}
         \caption{SAS}
         \label{fig:SAS_cl2_sil}
     \end{subfigure}
     \hfill
        \caption{Silhouette plot for the two implementations for 2 clusters}
        \label{fig:cl2_sil}
\end{figure}

\begin{figure}[htbp]
     \centering
     \begin{subfigure}[b]{0.48\textwidth}
         \centering
         \includegraphics[width=\textwidth]{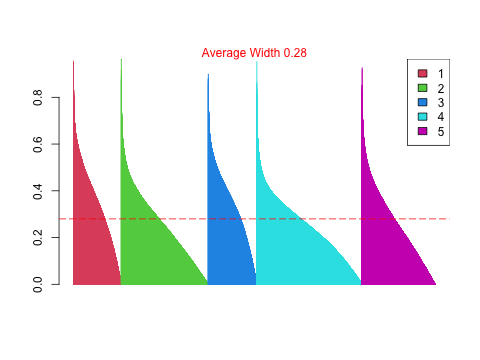}
         \caption{R}
         \label{fig:R_cl5_sil}
     \end{subfigure}
     \hfill
     \begin{subfigure}[b]{0.48\textwidth}
         \centering
         \includegraphics[width=\textwidth]{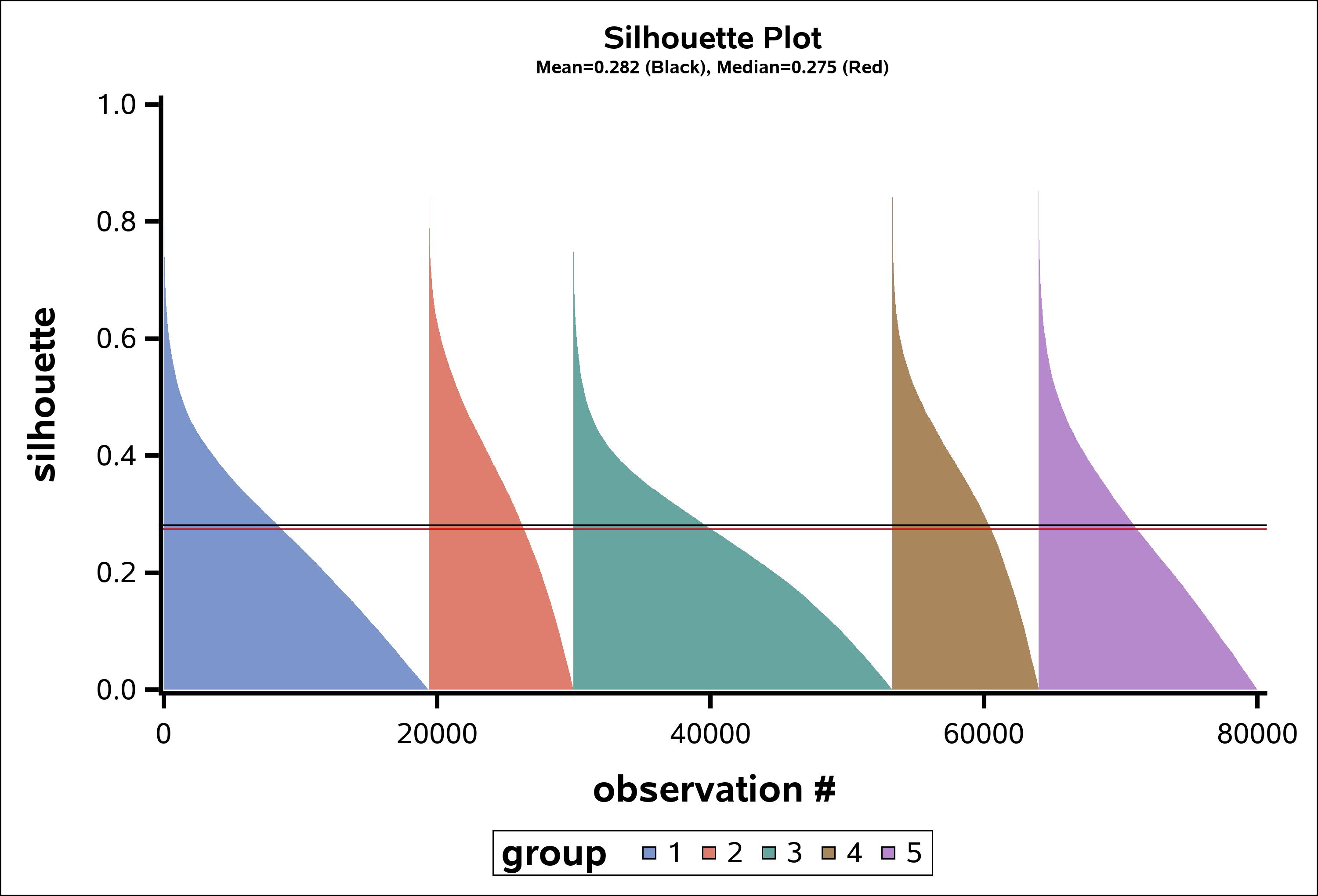}
         \caption{SAS}
         \label{fig:SAS_cl5_sil}
     \end{subfigure}
     \hfill
        \caption{Silhouette plot for the two implementations for 5 clusters}
        \label{fig:cl5_sil}
\end{figure}

\begin{figure}[htbp]
     \centering
     \begin{subfigure}[b]{0.48\textwidth}
         \centering
         \includegraphics[width=\textwidth]{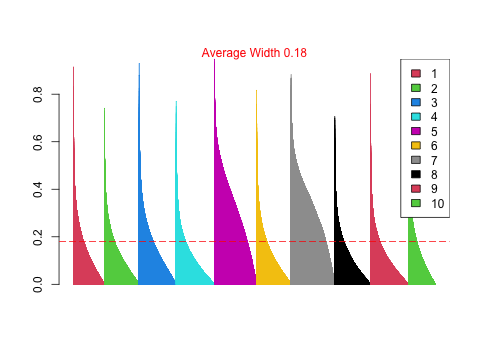}
         \caption{R}
         \label{fig:R_cl10_sil}
     \end{subfigure}
     \hfill
     \begin{subfigure}[b]{0.48\textwidth}
         \centering
         \includegraphics[width=\textwidth]{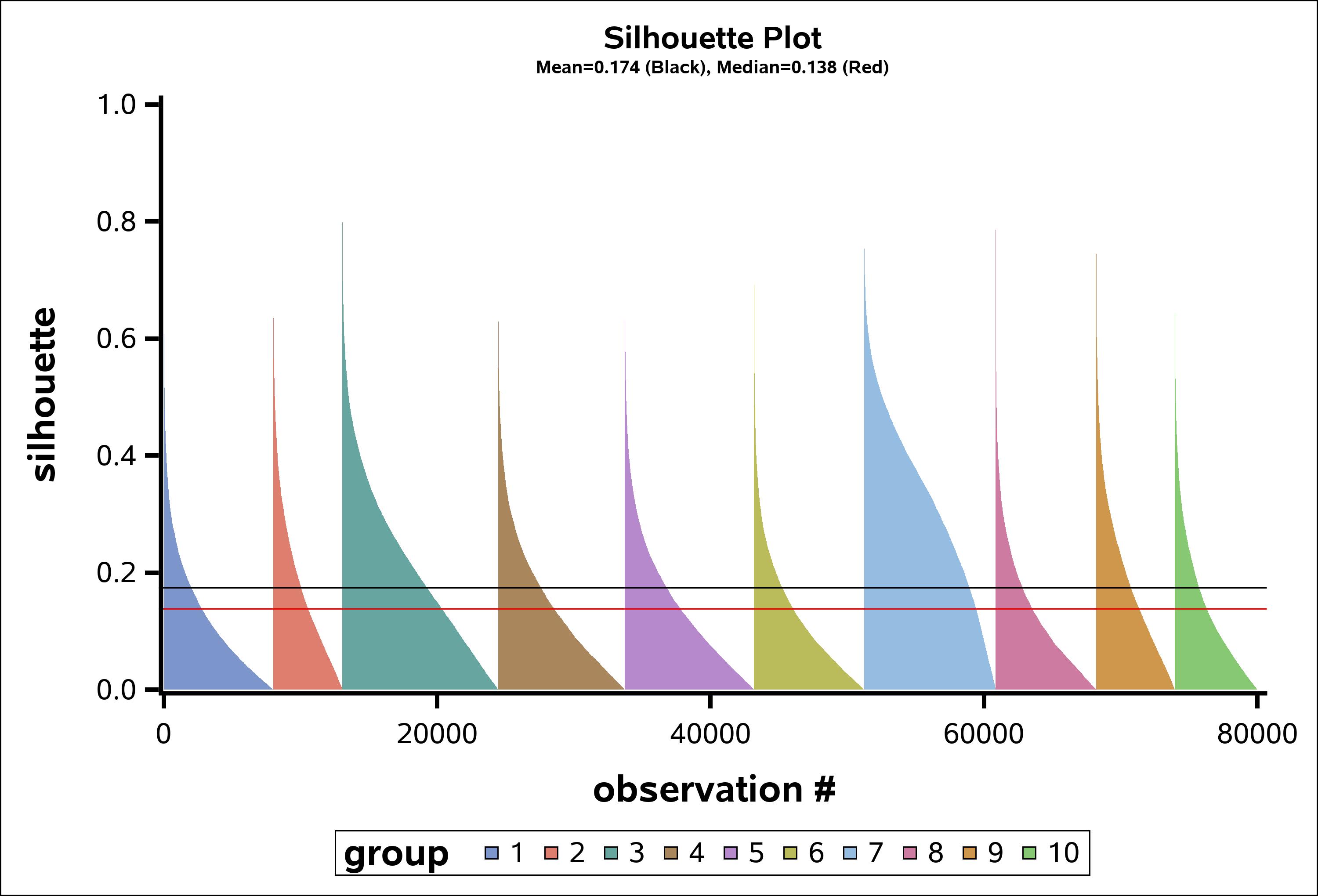}
         \caption{SAS}
         \label{fig:SAS_cl10_sil}
     \end{subfigure}
     \hfill
        \caption{Silhouette plot for the two implementations for 10 clusters}
        \label{fig:cl10_sil}
\end{figure}

Silhouette plots provide a way to select the number of clusters appropriate for our applications. They require distances between individual trajectories; however, this is not possible because of unequal trajectory sampling without fitting a separate model for each ID. Hence, the \code{clustra\_sil()} function uses trajectory distances to cluster mean spline trajectories instead. The \code{clustra\_sil()} function can take in either the output from a previous run of \code{clustra}, or can re-run the clustering with the specified parameters. Its output is a list that is used in conjunction with the \code{plot\_silhouette} function to generate silhouette plots. Each list element is a data frame with \code{cluster}, \code{neighbor}, and \code{silhouette} values. Below, in Figures \ref{fig:R_cl2_sil}, \ref{fig:R_cl5_sil}, and \ref{fig:R_cl10_sil},  we run it for the previously computed outputs of \code{clustra} with \code{k =} 2, 5 and 10.
\begin{verbatim}
sil = clustra_sil(cl5, mccores = mc,
    conv=c(20,0.5))
lapply(sil, plot_silhouette)

sil2 = clustra_sil(cl2, mccores = mc, 
    conv=c(20,0.5))
lapply(sil2, plot_silhouette)

sil10 = clustra_sil(cl10, mccores = mc, 
    conv=c(20,0.5))
lapply(sil10, plot_silhouette)

\end{verbatim}

Instead of using previous \code{clustra} output, the \code{clustra\_sil} function can run the clustering and generate the output from the original data by giving the  \code{clustra\_sil} function a dataset and a vector of cluster numbers instead of the output object from a \code{clustra} run. The following can also reproduce the figures.
\begin{verbatim}
sil = clustra_sil(data, k = c(2, 5, 10), 
    mccores = mc, conv = c(50, 0.5))
lapply(sil, plot_silhouette)
\end{verbatim}

\begin{figure}[htbp]
    \centering
    \includegraphics[scale=0.5]{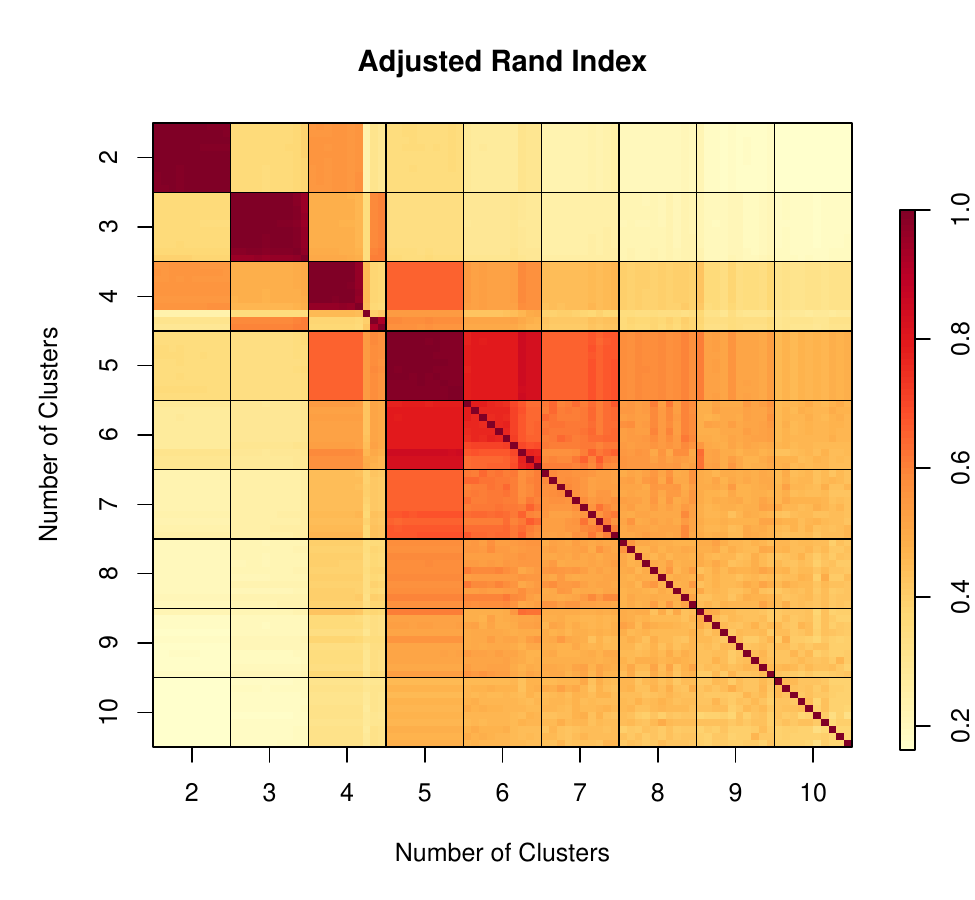}
    \caption{Adjusted Rand Index of cluster similarity for 10 random starts of 2, to 10 clusters based on data generated in 5 clusters.}
    \label{fig:clustra_rand}
\end{figure}

Another way to select the number of clusters is the Rand Index comparing different random starts and different numbers of clusters. When we replicate clustering with different random seeds, the "replicability" is an indicator of how stable the results are for a given $k$, the number of clusters. In this example, we look at \code{k = c(2,3,4,5,6,7,8,9,10)}, and 10 replicates for each $k$. The function \code{clustra\_rand()} will generate a dataframe with the adjusted rand index for all pairs from the trajectories results, which can then be fed into the \code{rand\_plot()} function to generate a matrix plot, given in Figure \ref{fig:clustra_rand}. 
\begin{verbatim}
ran = clustra_rand(data, k = seq(2, 10, 1), 
    starts = "random", mccores = mc, 
    replicates = 10, conv=c(40,0.5))
rand_plot(ran)
\end{verbatim}

The plot shows Adjusted Rand Index similarity level between all pairs of 90 clusterings (10 random starts for each of 2 to 10 clusters). The highest value of $k$ for which the ten random starts agree is 5. 

Another possible evaluation of optimum number of clusters is to first ask \code{clustra()} for a large number of clusters, evaluate the cluster centers on a common set of time points and feed the resulting matrix to a hierarchical clustering function. The code below asks for 40 clusters on our data and the function \code{hclust()} clusters the 40 cluster means, each evaluated on 100 time points. We also define a function \code{gpred} that generates the 100 predicted values for input to \code{hclust()}.
\begin{verbatim}
gpred = function(tps, newdata) {
  as.numeric(mgcv::predict.bam(tps, newdata, 
    type = "response", newdata.guaranteed = TRUE))
}
cl40 = clustra(data, k = 40, maxdf = 30,
    conv = c(100,0.5), mccores = mc)
timep = data.frame(time = seq(min(data$time), 
    max(data$time), length.out = 100))
resp = do.call(rbind, lapply(cl40$tps,
    gpred, newdata = timep))
plot(hclust(dist(resp)))
\end{verbatim}

\begin{figure}[tbp]
    \centering
    \includegraphics[scale=0.3]{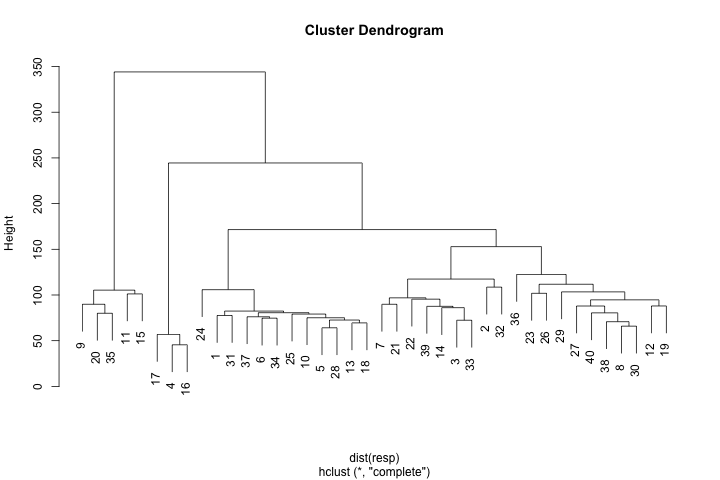}
    \caption{Cluster Dendogram for hierarchical clustering.}
    \label{fig:dendogram}
\end{figure}

The cluster dendrogram in Figure \ref{fig:dendogram} indicates there are five clusters if we draw the line at a height of about 140.

\subsubsection{SAS implementation of the data example} 

The \SAS macros can be included in the script as follows
\begin{verbatim}
 %include "\path\to\Macros\
    Trajectories Macros File v14R1.sas"
\end{verbatim}
We first initialize the graph settings. Details about each component can be found in the macro documentation, but the \code{\%graphset} macro governs the style of the output graphs from the macro
\begin{verbatim}
 %graphset(gpath=~\path, dpi=300, 
    style=splinecurve, format=jpeg,
    type=listing);
\end{verbatim}
The dataset can be read into \SAS as follows. 
\begin{verbatim}
proc import
    datafile = 'path\to\file\simul.csv'
    dbms=csv
    out=trajdata;
    getnames = yes;
    delimiter="09"x;
run;
\end{verbatim}
First we also initialize library to host the \SAS output.
\begin{verbatim}
libname traj5 "\path\to\library\";
\end{verbatim}
To run the trajectory clustering with 5 clusters we use the macro commands \code{\%trajsetup} followed by \code{\%trajloop}. 
\begin{verbatim}
%trajsetup(dsn=trajdata, id=id, time=time, 
    riskvar=response, ngroups=5, maxdf=30, 
    ptrim=0, seed=1256, steps=1, 
    method=0, random=YES);

%trajloop(outlib=traj5, iter=20, 
    minchange=0.5, minsubs=0, 
    min_x=-365, max_x=730, by_x=50,
    min_y=110, max_y=180, by_y=10, 
    showall=NO, showany=YES);
run;
\end{verbatim}
The \code{\%trajsetup} takes the dataset alongside the id, time and outcome variable as an input in the macro call. The macro also contains options to set the degrees of freedom, number of cluster groups as well as the methods options. Details about the methods available in the macro are provided in Appendix A, but for the purposes of this example we will use method=0, which uses \code{PROC GAMPL} for the spline estimation with the default optimization options.  The \code{\%trajloop} macro call will generate four \SAS datasets: \code{graphout}, \code{clusout}, \code{mergeit}, and \code{rms} in the specified library \code{traj5}, which contain results of the clustering. This macro also takes in parameters that control the x and y-axes of the resulting plots. We can use the \code{graphout} dataset to plot the trajectories. The \code{\%trajloop} macro will also generate a trajectory plot as part of it's default output, but we will use the following code for a cleaner plot.
\begin{verbatim}
proc sgplot data=traj5.graphout 
    /*noautolegend*/ dattrmap=graphattr;
    title 'Fitting 5 Clusters to the Data';
    series y=pred x=time / group=group attrid=scatter;
    xaxis  values=(-365 to 730 by 50)
        label=  "Time (in days from
            treatment initiation)";
    yaxis  values=(110 to 180 by 10) 
        label= "Predicted Systolic
        Blood Pressure";
run;
quit;
\end{verbatim}

Changing the \code{ngroups} option to 10, we get the results for 10 clusters.
\begin{verbatim}
libname traj10 "\path\to\library\";
%trajsetup(dsn=trajdata, id=id, time=time, 
    riskvar=response, ngroups=10, 
           maxdf=30, ptrim=0, seed=1256,
        steps=1, method=0, random=YES);
run;

ods select none;
%trajloop(outlib=traj10, iter=50, 
    minchange=0.5, minsubs=0,
    min_x=-365, max_x=730, by_x=50,
    min_y=110, max_y=180, by_y=10,
    showall=NO, showany=YES);
run;
proc sgplot data=traj10.graphout 
    noautolegend dattrmap=graphattr;
    title 'Fitting 10 Clusters to the Data';
    series y=pred x=time / 
        group=group attrid=scatter;
    xaxis  values=(-365 to 730 by 50) 
        label="Time (in days from 
            treatment initiation)";
    yaxis  values=(110 to 180 by 10) 
        label="Predicted Systolic
        Blood Pressure";
run;
quit;
\end{verbatim}

Similarly for 2 clusters we can change the \code{ngroups} option to 2 in \code{\%trajsetup} .
\begin{verbatim}
libname traj2 "\path\to\library\";
%trajsetup(dsn=trajdata, id=id, time=time, 
    riskvar=response, ngroups=2, 
    maxdf=30, ptrim=0, seed=1256, 
    steps=1, method=0, random=YES);

ods select none;
%trajloop(outlib=traj2, iter=20, 
    minchange=0.5, minsubs=0,
    min_x=-365, max_x=730, by_x=50,
    min_y=110, max_y=180, by_y=10,
    showall=NO, showany=YES);
run;
proc sgplot data=traj2.graphout
    noautolegend dattrmap=graphattr;
    title 'Fitting 10 Clusters to the Data';
    series y=pred x=time / 
        group=group attrid=scatter;
    xaxis  values=(-365 to 730 by 50) 
        label=  "Time (in days from
            treatment initiation)";
    yaxis  values=(110 to 180 by 10) 
        label= "Predicted Systolic
        Blood Pressure";
run;
quit;
\end{verbatim}
The output trajectories plots are given in Figures \ref{fig:SAS_cl2_sub}, \ref{fig:SAS_cl5_sub} and \ref{fig:SAS_cl10_sub}.
It is important to note that if the macro functions are used more than once in a single \SAS session, a new \code{\%trajsetup} macro call needs to be run every time before the \code{\%trajloop} macro is called, otherwise \SAS will use the options specified in a previous \code{\%trajsetup} matrix call.
The \code{\%silhouette} macro generates silhouette plots for cluster selection which are given in Figures \ref{fig:SAS_cl2_sil}, \ref{fig:SAS_cl5_sil} and \ref{fig:SAS_cl10_sil}, using the output generated by \code{\%trajloop}. If the libraries \code{traj2}, \code{traj5} and \code{traj10} contain the outputs for 2,5 and 10 clusters respectively, the we can generate the silhouette plots as follows.
\begin{verbatim}
%silhouette(lib=traj2, sid=id,groupsn=2, 
    cgroup=group);
%silhouette(lib=traj5, sid=id,groupsn=5,
    cgroup=group);
%silhouette(lib=traj10, sid=id,groupsn=10,
    cgroup=group);
\end{verbatim}

The output from these methods are available for reporting or further analysis. More details on the comparison between the SAS and the R implementations are provided in Appendix \ref{app:cluster_comp}.  

\section{Discussion}
Clustering of longitudinal trajectories has been a topic of interest across many disciplines. In the fields of biostatistics, epidemiology, and public health, there have been several approaches and computational algorithms proposed for classifying longitudinal data into groups. We present a series of \SAS macros and an \proglang{R} package, \pkg{clustra}, to assist with identifying patterns of longitudinal biomarkers over time that can be used to define population-based trajectory phenotypes. The purpose of this software is to facilitate the clustering of trajectory data in the hopes of discovering novel phenotypes that account for the totality of longitudinal observations collected on patients in EHR and large healthcare databases.

\cite{Teuling2023} evaluated several longitudinal clustering strategies using pre-existing \proglang{R} package and determined that two-step growth curve models with k-means and growth mixture models outperformed the k-means implementation from the \cite{Genolini2015} package. However, these methods were far more computationally intensive and applied to data with equal time intervals, thus may not be able to handle the large and irregular data like those of biomarkers in EHR. 

Like \cite{Luong2017}, our approach allows differing and irregular data for each subject, but we incorporate a smoothing spline rather than a fixed number of spline knots, allowing different levels of smoothness for each trajectory cluster. In particular, we chose thin plate splines as executed by the \SAS \pkg{GAMPL} procedure and \code{bam()} function of the \pkg{mgcv} \proglang{R} package for the smoothing component. These smoothing approaches are characterized by their built-in large-data capabilities, as well as, their use of a penalty to allow for automated smoothing with cross-validation. While our example focused on this smoothing approach, the macros and package include additional flexibility in choice of spline estimation method (see Appendix table \ref{app:macro}) and distance metric specification.

In this manuscript we offer some tools for determining and assessing the appropriate number of clusters in the K-means approach – silhouette plots, the Rand index, two-stage hierarchical clustering, etc. These methods provide guidelines for evaluating the appropriateness of the $K$ selected, however, we do not pretend to have solved the complex problem. For example, the Rand-plot (Figure \ref{fig:clustra_rand}) demonstrates that our 5-cluster assignment is more appropriate than a larger $K$ in our synthetic data, however, fewer clusters also appear to fit the data well. This is likely due to the simulated noise surrounding the ‘true’ clusters in the data, which was aimed at reflecting the variability seen in both the number of observations per patient and distances from the clustered-trajectory center in the real VA blood pressure data. With more well-defined curves, as in the clustra vignette \citep{clustra2024}, the algorithm is better able to discriminate between the groups.

To make this a truly dual-platform package, we have programmed both packages to act similarly and attempted to produce consistent results. Previous trajectory clustering packages like \pkg{kml3d} \citep{GENOLINI2013104} were developed only for \proglang{R}. Thus, a novelty of this approach is our development of parallel toolsets for two common programming languages, \proglang{R} and \SAS. Recently, there has been a push in other areas of computation to increase access to algorithms by implementing multi-platform software, such as \proglang{R} package \pkg{swdpwr} with corresponding SAS macros for power calculations \citep{CHEN2022106522} and \proglang{R} package \pkg{cccrm} with corresponding \SAS macros (\pkg{rm\_ccc())}) for estimating concordance correlation \citep{CARRASCO2013293}. Alternatively, some groups focus on providing code, tutorials, and implementation across several software platforms \citep{Smith2022,McGrath2020,gformula2021} for one software at a time. This is the first clustering approach to include strategies and code that work in both \proglang{R} and \SAS, relying on native functions and procedures already available in each computing environment. While some of the functionality may differ between the two software approaches, we have shown that the results are comparable using both (Appendix \ref{app:cluster_comp}). 

Methods have been proposed for multivariate clustering, such as deep learning methods using variational deep embedding with recurrence \citep{DEJONG2019} and the \proglang{R} package \pkg{clusterMLD} \citep{ZHOU2023} using hierarchical clustering, however both of these may be computationally inefficient on large data sets. Future directions for this work include the expansion of methods and code to the multivariate space, such as the joint modeling of systolic and diastolic blood pressures, or cholesterol measurements.

One important consideration within our algorithm is the specification of time zero for the data, prior to using the clustering functions. For the blood pressure example, we specifically selected individuals with no prior history of anti-hypertensive treatment and complete data on standard healthcare biomarkers and interactions in the year prior to treatment initiation, setting date of first medication fill as time zero. This allows for the interpretation of the clusters to align with the biological processes expected when someone initiates anti-hypertensive medication, and includes adjustment for the pre-initiation blood pressure measures that may have been critical in the decision to treat. Future work in this area should consider the portability of the predicted clusters, such that individuals may be classified to clinically-meaningful phenotypes outside of the clustered data. It is unclear whether these longitudinal trajectories are population-specific and how clinicians may use the predicted trajectories to identify intervention opportunities.

In conclusion, the \pkg{clustra} \proglang{R} package and corresponding \SAS macros integrate a K-means clustering algorithm for longitudinal trajectories that operates with large-scale EHR data. The corresponding clustering labels can be used as both exposure and outcome definitions when studying the relationships between health states. 

\bibliographystyle{SageH}
\bibliography{references.bib}
\newpage
\section{Appendix}
\subsection{SAS Macro Approaches}
\label{appendix:SASMacro}
The \SAS macros can use the \SAS procedures provided in the table below for creating cluster "centers" using splines. Method 0 was used in all the examples provided above.

\vspace{1em}
\label{app:macro}
    \begin{tabular}{|l|p{2in}|}
     \hline
  \bf Method 
  & \bf Description \\\hline
  0 & GAMPL with default optimization \\\hline
  1 & GAMPL with dual quasi-Newton optimization \\\hline
  2 & GAMPL with double-dogleg optimization \\\hline
  3 & GAMPL with conjugate-gradient optimization \\\hline
  4 & GAMPL with Nelder-Mead simplex optimization \\\hline
  5 & GAM \\\hline
  6 & GLIMMIX with EFFECT \\\hline
  7 & TPSPLINE \\\hline
  8 & TRANSREG \\\hline
\end{tabular}
\newpage
\subsection{Functional Comparisons Between SAS and R}

\vspace{1em}
\label{app:functions}
\begin{tabular}{|p{0.7in}|p{1.2in}|p{1in}|}
  \hline
  \bf Process 
  & \bf \proglang{R} functions 
  & \bf \SAS functions \\\hline
  Silhouette plot 
  & \code{clustra\_sil} 
  & Use \code{\%trajsetup}, \code{\%trajloop}, and \code{\%silhouette} respectively (or use wrapper \code{\%trajfit}) \\\hline
  Rand Index plot 
  & \code{clustra\_rand} 
  & Not available \\\hline
  Clustering at $k$ groups 
  & \code{clustra} 
  & \code{\%trajsetup} and \code{\%trajloop} respectively \\\hline
  Predicted values for $k$ 
  & Use \code{predict} function in \proglang{R} on the \code{tps} object created as an output by the function \code{clustra}
  &   “Graphout” dataset generated by \code{\%trajloop} \\\hline
  Plot of trajectories 
  & Use \code{plot\_smooths} on the data and the \code{tps} component of the \code{clustra} function output. 
  & \code{\%trajplot} (used after running \code{\%trajsetup} and \code{\%trajloop}) \\\hline
  Hierarchical clustering
  & A combination of fitting the \code{clustra} function on a large number of clusters and then plotting the predicted values using \code{hclust}
  & Use \code{\%trajsetup} with \code{STEPS=2} and a large number of clusters. Use the "WIDE" dataset created by \code{\%trajloop}as input to  \code{\%trajhclus} \\\hline
\end{tabular}

\newpage
\subsection{SAS vs. R Clustering Comparison} \label{app:cluster_comp}
The comparison of the results from the clustering for the \SAS macro as well as the \proglang{R} package is evaluated using a graphical check as well as the adjusted rand index. The comparison is performed in \proglang{R}, since the Adjusted Rand index can be calculated using the \code{MixSim::RandIndex} function. Additionally, the package \pkg{haven} is required to read in \SAS datasets. We first define a function \code{sas\_r\_cplot} to plot \SAS and \proglang{R} trajectories in the same graph. Since \SAS and \proglang{R} can produce different cluster labels, we need to map which clusters correspond to each other. The relabeling algorithm in \code{sas\_r\_cplot} matches pairs by distance, starting with shortest distance pair. The \code{file} parameter takes in the \code{graphout} file generated from a \SAS run and \code{clustra\_out} takes in the clustra run output. In the vignette, we also add a \code{sas.file()} function to construct the file path for the \SAS output on GitHub.
\begin{verbatim}
repo = "https://github.com/MVP-CHAMPION/"
repo_sas = paste0(repo, "clustra-SAS/raw/main/")
sas.file = function(file, rp = repo_sas) 
    paste0(rp, "sas_results/", file)
sas_r_cplot = function(file, clustra_out) {
  sas = cbind(haven::read_sas(sas.file(file)), 
    source = 1)
  r = data.frame(time = 
    min(sas$time):max(sas$time))
  n = nrow(r)
  k = nrow(sas)/n
  ## below assumes same time in each
  ## group and groups are contiguous.
  rpred = do.call(c,
  parallel::mclapply(clustra_out$tps,
    gpred, newdata = r, mc.cores = mc))

  ## compute mapping from SAS to R
  ## using manhattan distance over time
  rmap = vector("integer", k) 
  m = as.matrix(
    dist(rbind(t(matrix(sas$pred, n)), 
               t(matrix(rpred, nrow = n))), 
               "manhattan"))[1:k, (k + 1):(2*k)]
  for(i in 1:k) { 
    # assign starting with closest pair
    ind = arrayInd(which.min(m), dim(m))
    rmap[ind[1]] = ind[2]
    m[ind[1], ] = Inf
    m[, ind[2]] = Inf
  }
  
  r = cbind(r, group = sas$group, 
    pred = rpred, source = 2)
  plot(pred ~ time, data = sas, type = "n")
  for(i in 1:k) {
    sr = (1 + (i - 1)*n):(i*n)
    rr = (1 + (rmap[i] - 1)*n):(rmap[i]*n)
    lines(sas[sr, ]$time, sas[sr, ]$pred,
    lty = sas[sr, ]$source, col = i)
    lines(r[rr, ]$time, r[rr, ]$pred, 
    lty = r[rr, ]$source, col = i)
  }
  rmap # output the mapping from
       # SAS clusters to R: R = rmap[SAS]
}
\end{verbatim}
Using this function, we can generate plots with the \SAS (dashed line) and \proglang{R} (solid line) results overlayed on top of each other, with the same color indicating the corresponding clusters.

\begin{verbatim}
sas_r_cplot("graphout_cl2.sas7bdat", cl2)
sas_r_cplot("graphout_cl5.sas7bdat", cl5)
sas_r_cplot("graphout_cl10.sas7bdat", cl10)
\end{verbatim}
The adjusted rand index for the clustering comparison between the \SAS and \proglang{R} results can be obtained using the following code. It requires the \code{rms} dataset generated from the \SAS results alongside the \code{group} assignment generated by the \code{clustra} function.
\begin{verbatim}
MixSim::RandIndex(haven::read_sas
    (sas.file("rms_cl2.sas7bdat"))$newgroup, 
    cl2$group)
> 0.9963531
MixSim::RandIndex(haven::read_sas
    (sas.file("rms_cl5.sas7bdat"))$newgroup,
    cl5$group)
>0.9839461
MixSim::RandIndex(haven::read_sas
    (sas.file("rms_cl10.sas7bdat"))$newgroup,
    cl10$group)
> 0.3973877
\end{verbatim}
\begin{figure}[tbp]
    \centering
    \includegraphics[scale=0.5]{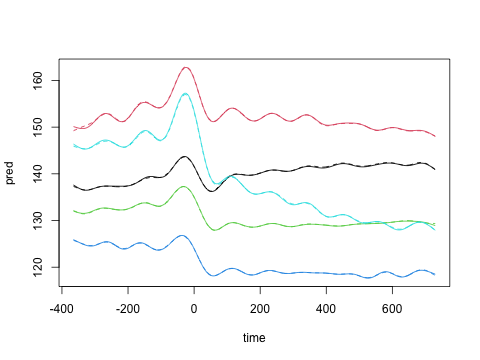}
    \caption{Comparing clustering from R (bold) and SAS (dashed) for 5 trajectories.}
    \label{fig:R_SAS_cl5}
\end{figure}
\begin{figure}[tbp]
    \centering
    \includegraphics[scale=0.5]{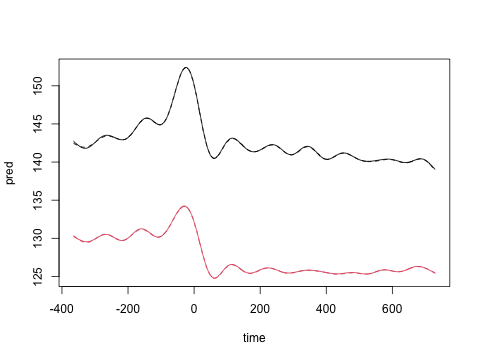}
    \caption{Comparing clustering from R (bold) and SAS (dashed) for 2 trajectories.}
    \label{fig:R_SAS_cl2}
\end{figure}
\begin{figure}[tbp]
    \centering
    \includegraphics[scale=0.5]{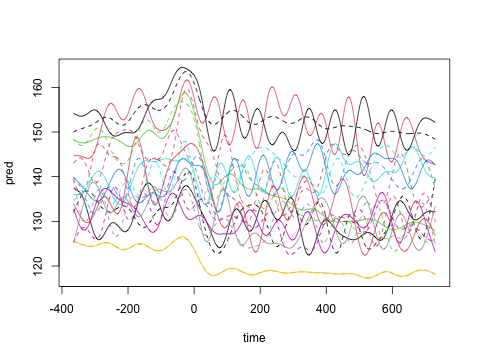}
    \caption{Comparing clustering from R (bold) and SAS (dashed) for 10 trajectories.}
    \label{fig:R_SAS_cl10}
\end{figure}

The adjusted rand indices for 2 and 5 clusters are very high, which indicate that there is great agreement in the classification of clustering between the \SAS and \proglang{R} results. The k=2 and k=5 cluster results are almost identical between \SAS (dashed lines) and R (solid lines). The pattern does not hold for the results from 10 clusters. However, it appears to keep the bottom cluster (yellow) from the 5-cluster result and is nearly identical between \SAS and \proglang{R}. The green clusters are also somewhat similar. The remaining 8 clusters seem to partition the data in a very unstable fashion that produces different results for different seeds. We know that the "true" cluster assignment is 5 clusters, and because of the over-specification of cluster centers plus the different random number generation methods in both \SAS and \proglang{R}, it is sensitive to noise in clustering assignments. Also, as we would expect, the Rand Index is not quite as strong. 
These results are also consistent with the full Rand Index plot in Figure~\ref{fig:clustra_rand}. In this example, the data was generated hence we have an idea of clustering performance with respect to the "true" data. In a real example, finding the true number of clusters can be a challenge. This should be done using underlying knowledge of the processes alongside statistical techniques as shown in this paper including using hierarchical modeling, the rand plot, etc. Clustering is dependent on random processes, we do not expect perfect agreement for both the macro and the package especially if over-specification is an issue, however both \SAS and \proglang{R} versions provide extremely similar utility for most cases.

\end{document}